\documentclass[pra,showpacs,twocolumn,superscriptaddress,amsmath,amssymb]{revtex4-1}
\pdfoutput=1

\usepackage{graphicx}
\usepackage{color}
\usepackage{dcolumn,amsmath}
\usepackage[normalem]{ulem}
\usepackage{xfrac}
\usepackage{epstopdf}

\begin{document}

\title{Laser controlled charge-transfer reaction at low temperatures}

\author{Alexander Petrov}
\affiliation{Department of Physics, Temple University, Philadelphia, Pennsylvania 19122, USA}
\affiliation{NRC Kurchatov Institute PNPI, Gatchina, Leningrad district 188300, Russia;
and Division of Quantum Mechanics, St.Petersburg State University, University Embankment 7-9,
St. Petersburg, 199034, Russia}

\author{Constantinos Makrides}
\author{Svetlana Kotochigova}
\affiliation{Department of Physics, Temple University, Philadelphia, Pennsylvania 19122, USA}

\email[Corresponding author: ]{skotoch@temple.edu}

\begin{abstract} 
We study the low-temperature charge transfer reaction between a neutral atom
and an ion under the influence of near-resonant laser light.   By setting up a 
multi-channel model with field-dressed states we  demonstrate that the reaction rate coefficient can be enhanced by
several orders of magnitude with laser intensities of $10^6$ W/cm$^2$ or larger. In addition, depending on laser frequency 
one can induce a significant enhancement or suppression of the charge-exchange rate coefficient.  For our intensities
multi-photon processes are not important.  
\end{abstract}


\maketitle

\section{Introduction}

The nature and mechanism of charge transfer reactions is of
interest to a wide range of scientific disciplines. For example, the ability to 
control  charge transfer processes is an important aspect of research on
solar cells \cite{YLi2015,Ponseca2016}, ion batteries 
\cite{Hausbrand2014,Choi2016}, ion sensors \cite{Shahinpoor2003,Cho2015}, 
and molecular electronics \cite{Kawai2009,Dell2015}. Various charge
exchange processes between particles in the universe serve as an important 
tool for  astrophysical research. The emitted photons are used to analyze and 
identify compositional and flux changes in  solar or stellar winds \cite{Lallement2004,Gu2016}.

Only recently  experimental techniques have become available allowing
the investigation of  charge transfer reactions between atoms and ions at
ultracold and cold temperatures
\cite{Grier2009,Zipkes2010,Schmid2010,Hall2011,Hall2012,Rellergert2011,Ratschbacher2012,Haze2015}.
These novel capabilities have paved the way towards explorations of the fundamental principles
of reactivity at the quantum level.  
Recent theoretical studies \cite{Cote2000,Watanabe2002,Zhao2004,Zhang2009,
Liu2009,Liu2010,Tacconi2011,Zhang2011,Lamb2012,Belyaev2012,Belyaev2013,McLaughlin2014}
mostly involve ultracold neutral alkali-metal or alkaline-earth atoms and fairly-cold alkaline-earth or 
rare-earth ions. This selection of atoms and ions has its background in the ready availability 
of these species in on-going experiments with hybrid atom-ion traps
\cite{Grier2009,Rellergert2011,Zipkes2010,Hall2011}.

Currently, the accuracy of these  theoretical studies is
limited by uncertainties in the short-range shape of the atom-ion
potentials, where the electronic clouds of the atoms
significantly overlap.  This state of affairs was not unlike that  for
neutral alkali-metal dimer potentials twenty five years ago when laser
cooling was at its infancy.  Clear exceptions are  systems dealing with
light atoms such as hydrogen~\cite{Bodo2008}.  In contrast the long-range
description of  atom-ion potentials is much better characterized.  The
leading term is an attractive $-C_4/R^4$ induction potential with a
coefficient $C_4$ that is proportional to the static polarizability of the
neutral atom and $R$ is the separation between the two particles.

In parallel, multi-channel quantum-defect theories have been developed that
parameterize the short-range atom-ion interactions in terms of boundary
conditions for the wavefunctions in conjunction with scattering from the
long-range potentials~\cite{Idziaszek2007,Idziaszek2009,Idziaszek2011,Li2012,Gao2013}.
They circumvent the need to explicitly know the short-range potentials by
replacing them with a few short-range parameters.  Reference~\cite{Li2012} observed that as few
as three parameters can describe the collisional dynamics over a relatively
large range of collisional energies $E$ extending from the ultra-cold regime to
about $E/k=1$ K, where $k$ is the Boltzmann constant.

Atom-ion charge transfer represents one of the simplest, yet still relevant
chemical reaction. Broadly speaking, charge transfer of distinct atomic species $A$ and $B$ can be classified by two processes:
non-radiative $A + B^+ \to A^+ + B$  and radiative $A + B^+ \to A^+ + B +\,\gamma$ 
or $A + B^+ \to AB^+ +\,\gamma$, where  a photon
$\gamma$ is emitted. In the last process, better known as radiative
association,  vibrationally-excited molecular ions are formed.
Radiative charge exchange with its spontaneously emitted photon is commonly
described with optical potentials~\cite{Zygelman1989,Liu2009,Liu2010}. 
Stimulated association due to the always-present black-body radiation can enhance the rates 
\cite{Zygelman1998}.

Several experimental and theoretical groups
\cite{Cote2000,Bodo2008,Grier2009,Li2012} have investigated 
charge-exchange collisions between different isotopes of the same atomic species, 
where the transition is nearly energetically resonant. The rate of such non-radiative processes approaches the Langevin rate.

An intriguing possibility in collisional physics and chemistry is that the reaction can be controlled
to achieve or optimize a specific outcome. 
Precedents for  manipulating charge-exchange reaction rates with radiation
exist in the literature. Reference~\cite{Vitlina1975} proposed to use a photon-induced 
crossing point between entrance-exit channels that occurs at a short internuclear separation,
whereas  Refs.~\cite{Tang1976,Tang1977} used a crossing point at large separation.
Their predictions for the reaction cross-section are based on the Landau-Zener theory 
of curve crossings. Later studies ~\cite{Hsu1985,Ho1985} made use of improved computational
capabilities to better describe the molecular electronic properties and thus
give a better understanding of controlled charge exchange. In addition, laser-induced control 
of chemical reactions is widely used in  chemistry in the thermal regime \cite{Brumer2003,Rice2000,Bandrauk1994,Ivanov2006}
and involves bound-to-continuum transitions in order to break  one or more bonds.
In the ultracold domain control might involve preparation of 
the initial reactants as used in \cite{Ratschbacher2012,Hall2012,Sullivan2012}. 

In this paper we explore  the control of the charge-exchange reaction at low temperature
by applying laser radiation with a frequency that is nearly
resonant to the energy difference between the entrance and exit channels.
Here, this implies that we study the continuum-to-continuum transition
$A + B^+ + n\gamma \to A^+ + B + (n+1)\gamma$, where $n$ is the photon number. 
We demonstrate that this stimulated radiative charge transfer can be enhanced by several orders of magnitude 
with laser intensities of $10^6$ W/cm$^2$ or larger. 
We focus on cold charge-exchange collisions between Ca and Yb$^+$ in the
presence of linearly-polarized radiation with wavenumbers between 400
cm$^{-1}$ and 1200 cm$^{-1}$ and intensities up to $10^{12}$ W/cm$^{2}$.
Transitions occur between  the excited A$^2\Sigma^+$ and ground X$^2\Sigma^+$
state potentials.

The paper is organized as follows. Section \ref{sec:adiab} reviews our
\textit{ab initio} adiabatic potential energy surfaces and describes our
procedure to diabatize the potentials. These diabatic potentials are then
used in Sec.~\ref{sec:cc} in a coupled-channels calculation that includes the
coupling to the radiation field.  Results on the laser intensity, collision energy, and
frequency dependence of the reaction rate are
given in Sec.~\ref{sec:results}.
We summarize the results in Sec.~\ref{summary}.

\section{Adiabatic and diabatic electronic potentials}\label{sec:adiab}

\begin{figure}
\includegraphics[scale=0.3,trim=0 0 0 0,clip]{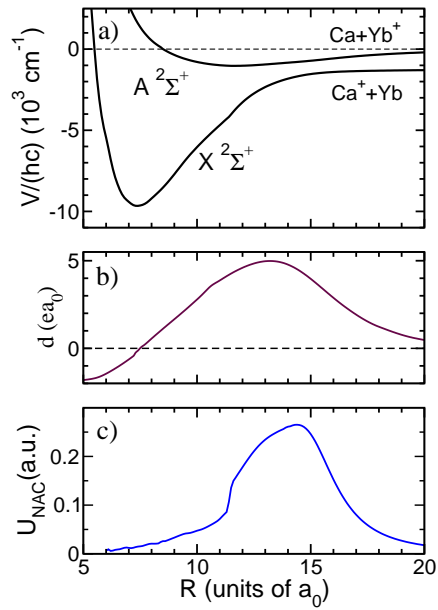}
\caption{Electronic properties of the CaYb$^+$ molecule. 
Panel a shows the ground X$^2\Sigma^+$ and excited 
A$^2\Sigma^+$  adiabatic potential curves as a function interatomic separation $R$. 
Panel b  and c show the transition electric dipole moment and the 
nonadiabatic coupling term $U_{\rm NAC}(R)$ between these two states as a function 
of $R$. For all panels lengths are expressed in the Bohr radius $a_0=0.0529$ nm
and the dipole moment is given in units of $ea_0$, where $e$ is the charge of the electron.
}
\label{structure}
\end{figure}

We, first, determine the non-relativistic ground and excited CaYb$^+$
potentials as a function of internuclear separation $R$ using multi-configuration 
second-order perturbation theory (CASPT2) implemented in the
MOLCAS software suite \cite{Karlstrom2003}.  
Reference wave functions are
obtained from a complete active space self consistent field (CASSCF)
calculation with 4s4p5s orbitals of Ca and  6s6p7s orbitals of Yb in the
active space.  This is followed by a CASPT2 calculation, where the 3s$^2$ 3p$^6$
electrons of Ca and  5s$^2$ 5p$^6$ 4f$^{14}$ electrons of Yb are
correlated. Here, the TZVP ANO-RCC (triple-zeta valence polarized 
atomic natural orbital relativistic CASSCF/CASPT2) basis sets \cite{manual} 
contain  (20s 16p 6d 4f ) [6s 5p 2d 1f] functions for Ca and (25s 22p 15d 11f 4g 2h) [8s 7p 5d 3f 2g 1h]
functions for Yb. The relevant electronic  dipole moments and non-adiabatic coupling term
between the CaYb$^+$ and Ca$^+$Yb molecular potentials  have also been 
calculated using the CASSCF method.

The two energetically-lowest potentials, $V_X(R)$ and  $V_A(R)$, for the $^2\Sigma^+$ symmetry are presented in
Fig.~\ref{structure}a.  We see that the ground X potential  
is much deeper than the excited A potential and that they have an avoided crossing at 
$R_c=15a_0$ with a splitting of $\Delta V/(hc)=1000$ cm$^{-1}$, where $h$ is the Planck constant and $c$ is the speed of light.
These potentials dissociate to the atomic Ca$^+$($^2$S) + Yb($^1$S) and Ca($^1$S) + Yb$^+$($^2$S) limits,
respectively. The two limits are split by $\Delta/(hc)=1136$ cm$^{-1}$.
The potentials
have an attractive long-range $-C_4/R^4$ tail, where $C_4 = 71.5 E_ha_0^4$ for the X state \cite{Dalgarno2007} and $78.5 E_ha_0^4$ for the A state \cite{Derevianko2006}. 
Here $E_h$ is the Hartree energy.  Figure~\ref{structure}b shows the electronic transition dipole
moment $d(R)$ between these $^2\Sigma^+$ states. The transition dipole moment has a maximum at $R=13
a_0$ and approaches  zero for large interatomic separations.

Other excited state potentials (not shown in Fig.~\ref{structure}a 
lie $\approx$13000 cm$^{-1}$ above the A potential and do not contribute to the reaction,
so that in our computation we can focus on these two lowest electronic potentials.

\begin{figure*}
\includegraphics[scale=0.3,trim=0 0 0 0,clip]{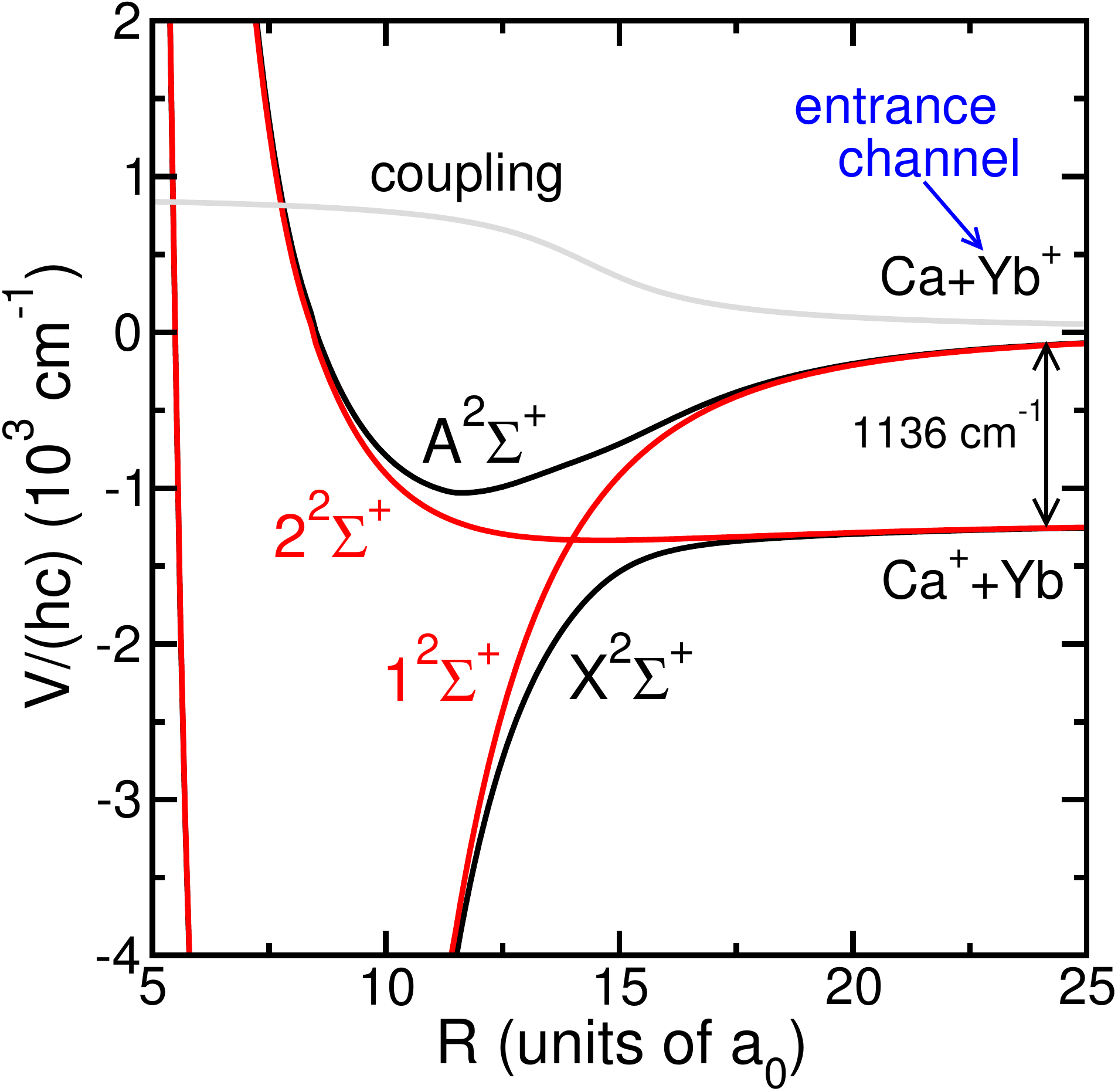}
\includegraphics[scale=0.3,trim=0 0 0 0,clip]{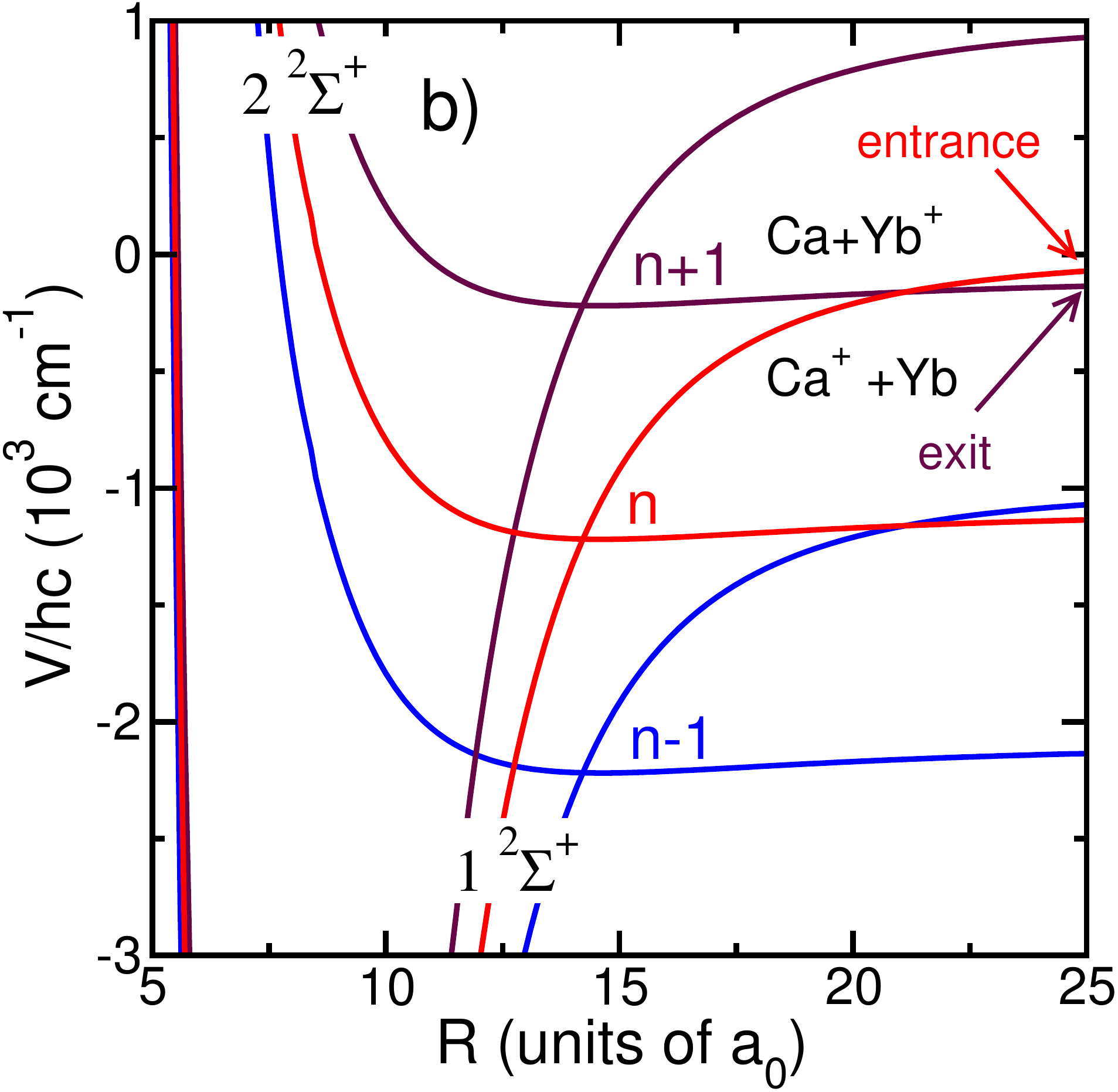}
\caption{Panel a: The ground X$^2\Sigma^+$ and excited A$^2\Sigma^+$ adiabatic
(black curves) and the 1$^2\Sigma^+$ and  2$^2\Sigma^+$ diabatic (red curves) potential 
energy curves of  CaYb$^+$  as a function of $R$. Only energies near the dissociation limits are shown.
Panel b: Dressed picture of the CaYb$^+$ potentials  as a function of $R$ with $n-1$, $n$, and $n+1$ photons and $\ell=0$. 
The photon wavenumber is 1000 cm$^{-1}$.
Here, we assume that the entrance channel corresponds to the potential with $n$ photons that asymptotically has the largest energy. }
\label{pots}
\end{figure*}

The corresponding electronic molecular states, denoted by $|X;R\rangle$ and
$|A;R\rangle$, parametrically depend on $R$. Consequently, 
 non-adiabatic coupling between the potentials, proportional to
$U_{\rm NAC}(R)=\left< X;R \right| d/dR \left| A;R \right>$ and shown in Fig.~\ref{pots}c, leads to 
non-radiative charge transfer reaction when the collision entrance
channel is Ca($^1$S) + Yb$^+$($^2$S), the continuum of the A state potential.  

In order to set up the closed-coupling calculation we  diabatize the adiabatic X and A potentials
by introducing the two  diabatic wavefunctions 
\begin{equation}
 \left(\begin{array}{c}
|1 \rangle \\
|2 \rangle
\end{array}\right)
=
  O(R) \left(\begin{array}{c}
|X;R\rangle\\
|A;R\rangle
\end{array}\right)
\end{equation}
with orthogonal transformation
\begin{equation}
O(R)= \left(
\begin{array}{cc}
\cos(\theta) & \sin(\theta) \\
-\sin(\theta) & \cos(\theta)
\end{array}\right)
\end{equation}
and  angle $\theta(R) = \int_{R}^{\infty}U_{\rm NAC}(R')dR'$.

Diabatic states $\left|1 \right>$ and $\left|2 \right>$, by {\it construction} are assumed to be $R$ independent, are coupled according to the $2\times2$
potential matrix
\begin{eqnarray}
   V^{\rm mol}(R)&=& \left ( 
        \begin{array}{cc}
            V_1(R)  &  V_{12}(R)\\
            V_{12}(R) & V_2(R)
        \end{array}
        \right)  \nonumber \\
       &=& O(R)
\left ( \begin{array}{cc}
V_X(R) & 0 \\
0 & V_A(R)
\end{array}  \right)
O^T(R)\,,
\end{eqnarray}
where $O^T(R)$ is the matrix transpose of $O(R)$.
The diabatic potentials $V_1(R)$ and $V_2(R)$ and coupling function $V_{12}(R)$ are shown
in Fig.~\ref{pots}a. 
Similarly, the dipole moment matrix in the diabatic basis set is 
\begin{equation}
\left ( 
        \begin{array}{cc}
D_1(R) & D_{12}(R) \\
D_{12}(R) & D_2(R)
\end{array}
        \right)
=
O(R)
\left ( 
        \begin{array}{cc}
d_X(R) & d(R) \\
d(R) & d_A(R)
\end{array}
        \right)
O^T(R)\,,
\label{Ddiab}
\end{equation}
where $d_X(R)$ and $d_A(R)$ are the permanent dipole moments of the X and A states.

\section{Coupled-channels calculation}\label{sec:cc}

We set up a multi-channel description of charge-exchange
collisions between an atom and an ion in the presence of near-resonant linearly-polarized laser light
using the dressed-state picture or the Floquet Ansatz. We construct coupled
time-independent radial Schr\"odinger equations for the interatomic separation
$R$ in basis states $|\alpha, m_S; \ell m_\ell;n\rangle= |\alpha, m_S\rangle |\ell m_\ell\rangle |n\rangle$, 
where $ |\alpha, m_S\rangle$  are the diabatic $^2\Sigma^+$ electronic states with $\alpha=1$ or $2$ and 
$m_S=\pm 1/2$ is the projection quantum number of the electron spin $S$ of
the $^2$S ion along the laser polarization. The ket $|\ell m_\ell\rangle \equiv Y_{\ell m_\ell}(\hat R)$ 
is a spherical harmonic describing the relative rotational wavefunction of the two particles
around the center of mass with projection quantum number $m_\ell$ along
the direction of the laser polarization. Finally, $|n\rangle$ is a Fock state with $n$  laser 
photons of frequency $\omega$. 

Figure~\ref{pots}b illustrates the dressed states picture of the CaYb$^+$ potentials with $n-1$, $n$, and $n+1$ photons and $\ell=0$. 
The figure shows three pairs of diabatic potential curves shifted by photon energy $\hbar\omega$ with $\hbar=h/(2\pi)$. For the 1000 cm$^{-1}$
laser frequency, used in the figure, an additional crossing near $R=21a_0$ is created between the entrance channel with $n$ 
photons and an exit channel with $n+1$ photons. 
This leads to a new reaction pathway with an exit channel with a small relative kinetic energy between the particles.

\begin{figure*}
\begin{center}
\includegraphics[scale=0.33,trim=0 0 0 0,clip]{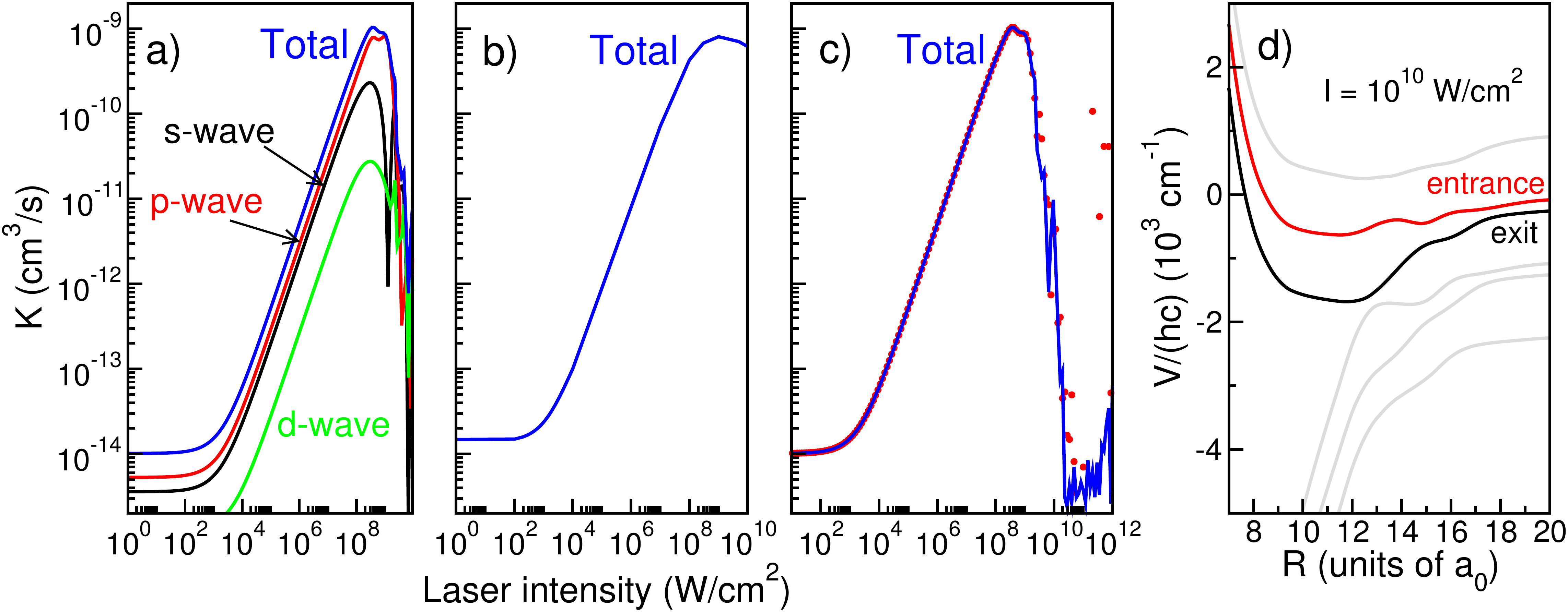}
\end{center}
\caption{Charge-exchange rate coefficient for Ca + Yb$^+$ as a function of laser intensity
over 10 decades and a laser wavenumber of 1000 cm$^{-1}$. Panel a: The collision
energy is $E/k=1$ $\mu$K with one-photon dressing $\delta n=1$. The contribution to the rate of several partial waves is
shown by different colored curves. Panel b: The collision energy is $E/k=1$ mK with the same 
one-photon dressing $\delta n=1$. 
Panel c: The total rate coefficient at collision energy of $E/k=1$ $\mu$K is shown by a 
solid blue line for one-photon dressing and by red markers for two-photon dressing. 
Panel d: The adiabatic field-dressed potential surfaces that are Stark-shifted and modified by a strong
laser field at intensity of 10$^{10}$ W/cm$^2$. The laser frequency is 1000 cm$^{-1}$ and E/k = 1$\mu$K.}
\label{intensity}
\end{figure*}

The Hamiltonian of our system is
\begin{equation}
H =-\frac{\hbar^2}{2\mu_r}\frac{d^2}{dR^2} + \frac{\bf L^2}{2\mu_rR^2} + 
V^{\rm mol}(R) + V^{\rm rad}({\bf R}) + \hbar \omega a^\dagger a,
\label{eq:ham}
\end{equation}
where $\mu_r$ is the reduced mass, ${\bf L}$ is the rotational angular momentum operator, 
$|\ell m_\ell\rangle$ are eigenstates of ${\bf L^2}$, and $V^{\rm mol}(R)$ is
electronic Hamiltonian defined in the previous section.  For a $^2\Sigma^+$
system the Hamiltonian $V^{\rm mol}(R)$ is isotropic and does not affect or
couple rotational states.  

The last two terms in Eq.~(\ref{eq:ham}) describe the coupling between the
particles and laser field and  the Hamiltonian of the field, respectively.
In the dipole and long-wavelength approximations the molecule-field
interaction $V^{\rm rad}({\bf R})= -\sqrt{2 \pi \hbar \omega/V}\,
(\vec{\epsilon} \cdot \vec{D}) (a^\dagger + a)$, where $\vec D$ is the molecular electric dipole
moment operator, constructed from Eq.~\ref{Ddiab}, and the field operators $a^\dagger$ and $a$ create and
destroy laser photons of frequency $\omega$ and polarization $\vec \epsilon$
in volume $V$.  This molecule-field interaction  is
anisotropic and only has non-zero matrix elements between states that differ
by one photon.  We choose the polarization vector $\vec{\epsilon}$ along the
laboratory $z$ axis. In our basis the matrix elements are
\begin{eqnarray}
\lefteqn{ \langle 1, m_S;\ell m_\ell;n+1|V^{\rm rad}|2, m_S; \ell'm'_\ell;n\rangle} \\
   &=& -D_{12}(R)\sqrt{\frac{2 \pi I} {c}} \sqrt{\frac{2\ell'+1}{2\ell+1}}C^{\ell m_\ell}_{10,\ell'm'_\ell}C^{\ell0}_{10,\ell'0},
              \nonumber
\end{eqnarray}
where $D_{12}(R)=\langle 1, m_S| d_z | 2, m_S\rangle $ is the electronic transition dipole moment for our linearly polarized photon,  and
$I=nc\hbar\omega/V$ is the laser intensity. The
functions $C^{jm}_{j_1m_1,j_2m_2}$ are Clebsch-Gordan coefficients.

The charge exchange rate coefficient from $|1,m_S\rangle$ with $n$ photons is given by
\begin{equation}
K = \frac{\hbar \pi}{\mu_r k} \sum_{\ell,\ell'=0}^{\ell_{\rm max}}\sum_{m_{\ell}}\sum_{n'=n-\delta n}^{n+\delta n}  \left| T_{2,\ell' m_{\ell},n'\leftarrow1,\ell,m_{\ell}, n} \right|^2,
\label{sum}
\end{equation}
where $k$  is collisional wave vector, $m_{\ell}$ varies from $-{\rm min}(\ell, \ell')$ to
${\rm min}(\ell, \ell')$, $\delta n$=1 or 2 in our simulations, and $T_{f\leftarrow i}$  are 
T-matrix elements obtained from the scattering solutions
of Eq.~\ref{eq:ham}. We find that the main contribution to $K$ comes from
$T$-matrix elements with $n'™=n+1$, which corresponds to the transition
 $|1\rangle +  n\hbar\omega \rightarrow |2\rangle +  (n+1)\hbar\omega$. 

We neglect effects of the permanent dipole moments $D_1(R)$ and $D_2(R)$ as the detuning between
states of the same molecular state but with different photon number is
large. Moreover, such coupling does not lead to charge exchange.

To calculate the non-radiative and stimulated radiative charge exchange rate we include many partial
waves for both continua.  For example, $E/k$ = 1 mK and 10 mK requires
$\ell_{\rm max}\approx$ 10 and 20 partial waves, respectively.  In contrast, we only need to
include a few partial waves (up to 3 or 4) for  collision energies around 1
$\mu K$.

\section{Results}\label{sec:results}

We modify the charge-exchange reaction by applying an infra-red laser
with frequency near 1000~cm$^{-1}$.  This is a natural choice of the frequency, which
is resonant to the adiabatic potential splitting near the avoided crossing at $R=R_c$. 
Figures~\ref{intensity}a and b show the charge-exchange rate coefficient as a
function of laser intensity for collision energies of $E/k=1$ $\mu$K and 1 mK, respectively.  
In both panels the intensity is varied over 10 decades, leading to a dramatic change 
of the rates coefficient. 

At low laser intensity, from zero up to $10^{2}$ W/cm$^2$, the weak non-adiabatic 
interaction between the adiabatic X and A potentials results in a small reaction rate 
of the order of $10^{-14}$ cm$^3$/s.  As the field intensity is increased, from 
$10^{3}$ to $10^{8}$ W/cm$^2$, we observe a
linear rise in the reaction rate reaching a maximum value close to the Langevin rate
for our system. Figure~\ref{intensity}a also shows the partial wave contribution to the
reaction rate coefficient for $E/k=1$ $\mu$K.  Only $s$, $p$, and $d$ waves
contribute significantly.  The calculation becomes more complex for a
collision energy of 1 mK in Fig.~\ref{intensity}b as a greater number of
channels  are coupled by the laser field. We obtain the total charge-
exchange rate coefficient $K$ after summation of the transition matrix
elements squared (Eq.~\ref{sum}) over all possible m$_{\ell}$ projections
for  partial waves from 0 to $\ell_{\rm max}$=10. For clarity contributions
from individual partial waves are not shown.

The effect of one- or two-photon dressing (i.e. $\delta n=1$ or $2$) is also
studied  in Fig.~\ref{intensity}c.  The comparison between these two cases
demonstrates that the one-photon dressing model is adequate for our hybrid
system when the intensity is below 10$^{10}$ W/cm$^2$. These results are
observed for a collision energy of 1 $\mu$K.  Finally, we
note that the total reaction rates in Figs.~\ref{intensity}a and b obtained
for different collision energies are nearly the same despite of large
difference in the contributing number of partial waves.

The strong resonant field creates coupling between 
the entrance and light-induced exit channels, indicated in Fig.~\ref{pots}b, and leads to
an enhanced reaction rate. Even stronger laser fields with an intensity above 
$10^{9}$ W/cm$^2$ modifies energies of the interaction potentials inducing a
reaction barrier and thereby slowing the reaction. We demonstrate this effect by showing
 Stark-shifted adiabatic potentials in Fig.~\ref{intensity}d at $I=10^{10}$ W/cm$^2$ and laser
frequency of 1000 cm$^{-1}$. 

\begin{figure*}
\includegraphics[scale=0.35,trim=0 0 0 0,clip]{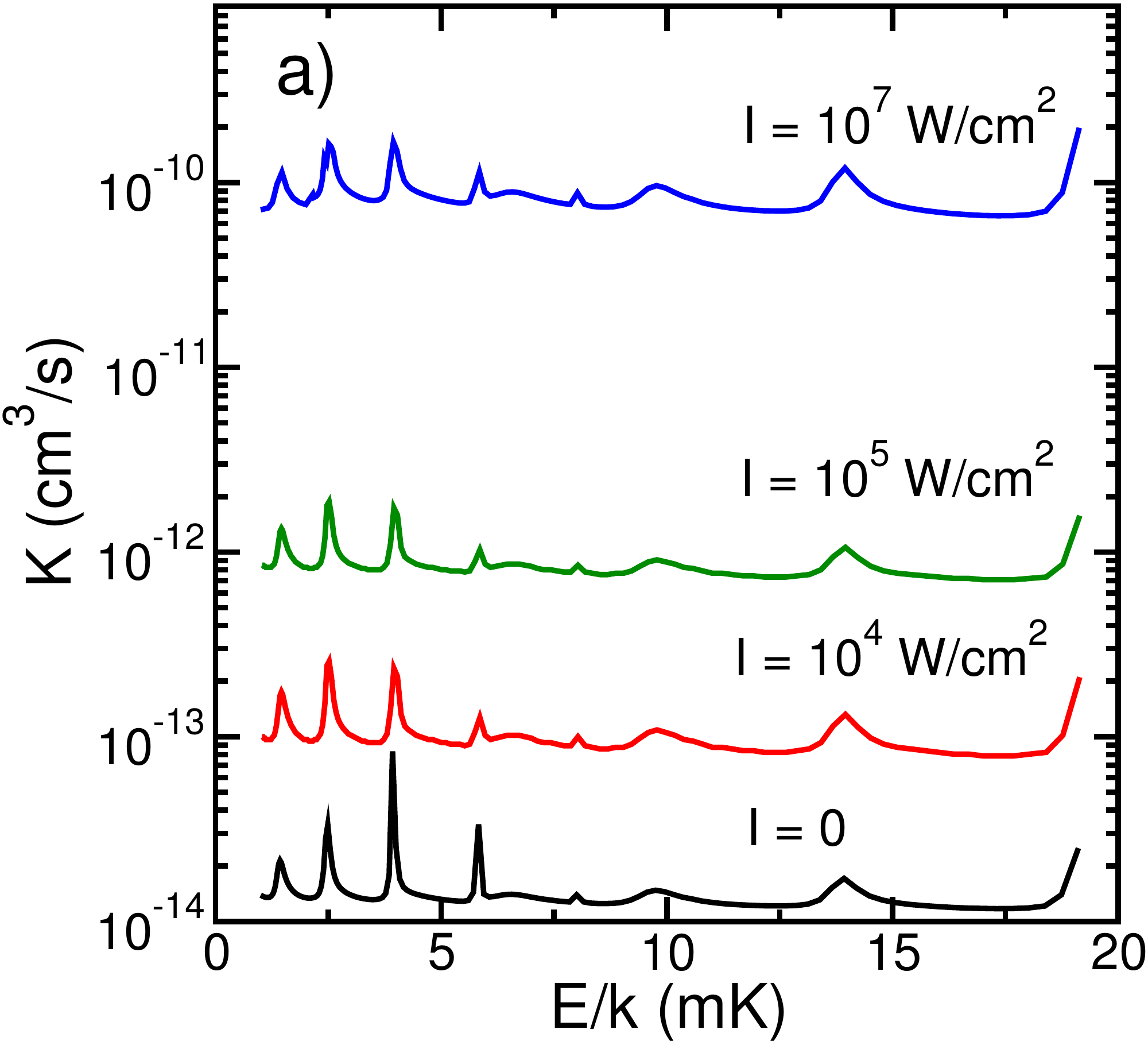}
\includegraphics[scale=0.35,trim=0 0 0 0,clip]{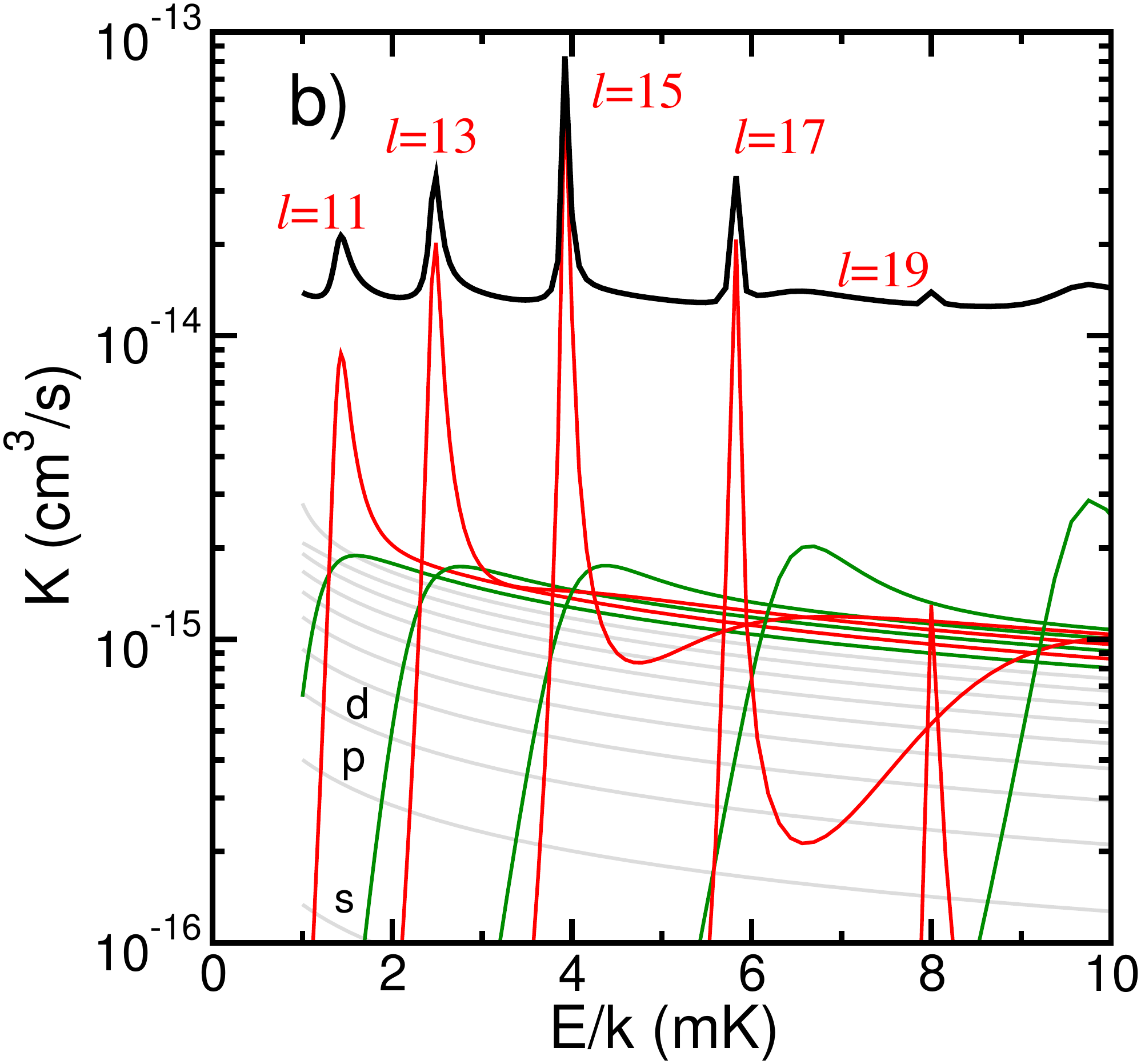}
\caption{Panel a: Charge-exchange rate coefficient for Ca + Yb$^+$ as a function of collision energy 
E/k from 1 mK  to 20 mK for four laser intensities. Panel b:  The partial wave contributions to
the rate coefficients at zero intensity. The gray curves correspond to partial rates for $\ell \leq 9$, 
those for $s$, $p$, and $d$ waves are indicated in the figure. From left to right the dark-green curves  
correspond to  $\ell = 10, 12, 14, 16$, and 18, respectively. Partial rates for $\ell = 11, 13, 15, 17$, and $19$,
shown by red curves, have distinct orbiting resonances.
The wavenumber of the laser is 1000 cm$^{-1}$  and $\delta n$ = 1 in both panels.}
\label{partial_waves}
\end{figure*}

The total charge-exchange rate coefficient $K$ as a function of collision
energy is shown in Fig.~\ref{partial_waves}a for several intensities.  
In the absence of the external field the rate coefficient is relatively small, only slightly above
10$^{-14}$ cm$^3$/s, and increases four orders of magnitude when the intensity
reaches 10$^{7}$ W/cm$^2$. This occurs  uniformly over the 20 mK interval of collision energies
shown in the figure.
Moreover, the unthermalized rate coefficient has a number
of shape resonances due to the many partial waves that contribute.   The resonances
occur when the energy of the entrance channel matches quasi-bound levels
trapped by the long-range potential, $-C_4/R^4 + \hbar^2 \ell (\ell + 1)/(2\mu_r R^2)$, near the top of the centrifugal barrier. This
enhances the wave functions at small separations and thus enhances the
charge-exchange rate coefficient.  We also find that the resonance positions
occur at almost the same collision energy at any laser intensity. There is
a slight shift and broadening of the resonances when the intensity is increased.
Once we thermalize the rate coefficient the effect of resonances is less visible.

In our calculations we are able to identify the partial waves that contribute
to the resonances. Figure~\ref{partial_waves}b presents plots of 
partial rate coefficients at zero field intensity for $\ell$ = 0 to 19
as a function of collision energy. For collision energies from 1 mK to 10 mK resonances occur for odd partial-wave 
quantum numbers between 11 and 19.  Reference \cite{Gao2010} showed 
that the analytical scattering solutions for a $-C_4/R^4$ potential are such that if
a shape resonance exists for partial wave $\ell$ then  $\ell+2, \ell+4$ etc. resonances
also exist. This observation explains our finding that all resonances in Figs.~\ref{partial_waves}a and b are due to odd partial waves.

\begin{figure}
\includegraphics[scale=0.33,trim=0 0 0 0,clip]{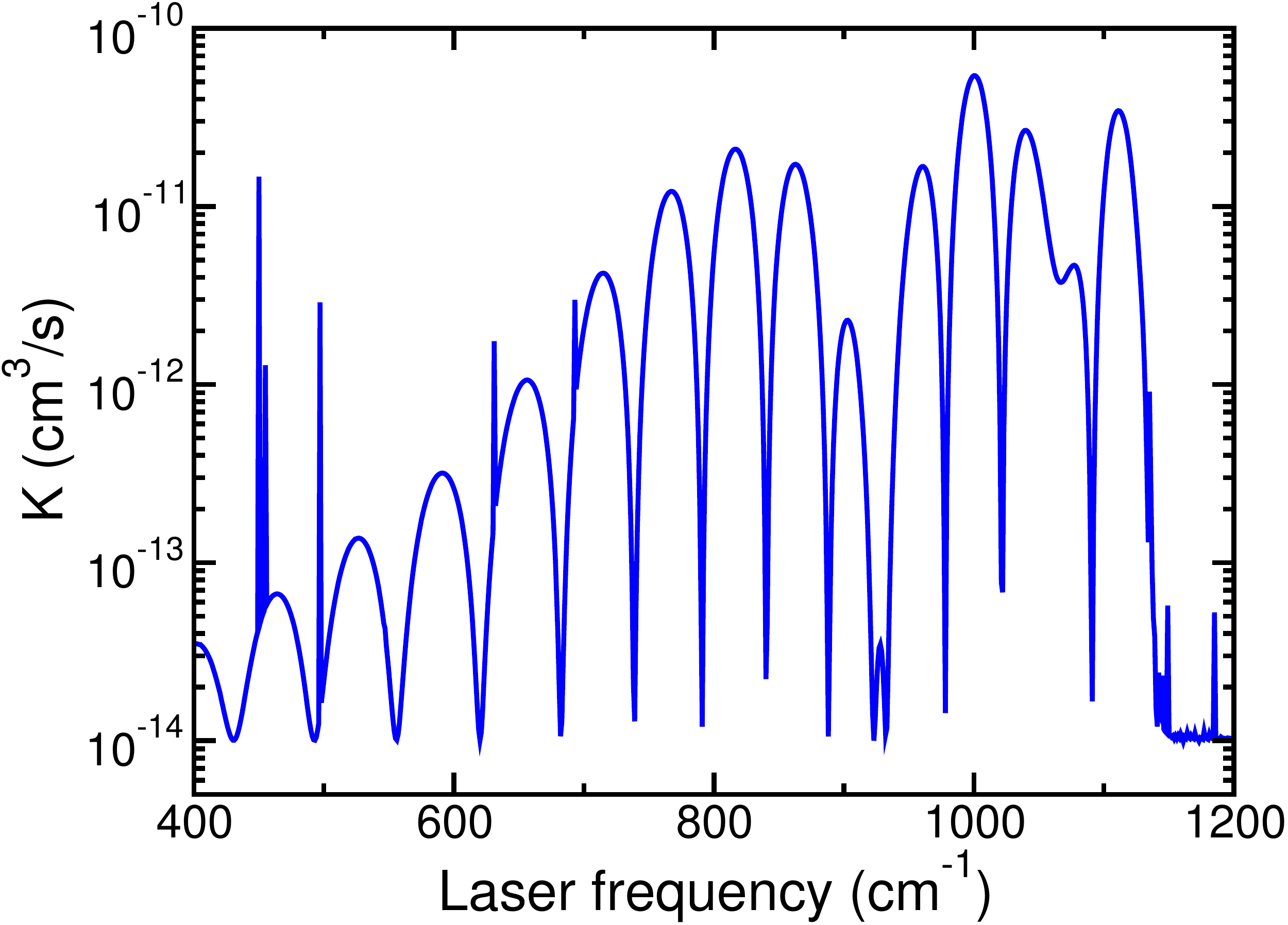}
\caption{Charge-exchange rate coefficient  for Ca + Yb$^+$ as a function of laser wavenumber.
Laser intensity is $10^7$ W/cm$^2$ and $E/k=1 \mu$K. }
\label{frequency}
\end{figure}

We also test the dependence of the charge-exchange rate on the laser
wavenumber in between 400 cm$^{-1}$ and 1200  cm$^{-1}$.
Figure~\ref{frequency} shows that the rate coefficient is very sensitive to
the laser frequency and even oscillates due to changing wave-function
overlap at the avoided crossing point between the initial $n$ photon state and the final $n+1$ photon state. 
The location of this crossing point also changes with the laser frequency.
In fact, these are Stueckelberg oscillations. The maximum rate is reached for a wavenumber
$\omega/c\approx1000$ cm$^{-1}$, which corresponds to the closest approach of the
adiabatic potentials.  For $\omega/c>1136$ cm$^{-1}$, the energy difference
between the dissociation limits, the laser light does not affect the charge-exchange rate and
leads to $K$ of the order 10$^{-14}$ cm$^3$/s. The laser-induced crossing between entrance and exit channels
shown in Fig.~\ref{pots} does not occur any more.

\section{Summary}
\label{summary}
We have explored the effect of a moderately intense laser field on the charge transfer
reaction for hybrid atom-ion collisions in the realm of cold temperatures. 
We have shown that the reaction rate coefficient can be significantly enhanced 
with a near-resonant laser field.
We find that around a field intensity of 10$^3$ W/cm$^2$ the reaction mechanism 
changes from being dominated by the intra-molecular non-adiabatic coupling to
being laser field dominated.
We investigate these processes over a
wide range of laser intensities, laser frequencies and collision energies.

\section{Acknowledgments} This work is supported by grants of the ARO-MURI Nos. 
W911NF-14-1-0378 and W911NF-12-1-0476, the NSF No. PHY-1619788. 

\bibliography{ChargeExchange}

\end{document}